\documentclass[pra,a4paper,twocolumn,a4paper,showpacs,amsmath,amssymb]{revtex4}

\usepackage{graphicx}% Include figure files
\usepackage{dcolumn}% Align table columns on decimal point
\usepackage{bm}% bold math
\begin{document}

\preprint{APS/123-QED}

\title{Stability of the self-phase-locked pump-enhanced singly resonant parametric oscillator}% Force line breaks with \\

\author{Jean-Jacques Zondy}
 \email{jean-jacques.zondy@obspm.fr}
\affiliation{BNM-SYRTE (UMR-CNRS 8630), 61 avenue de l'Observatoire, F-75014 Paris, France}%

\date{\today}% It is always \today, today,

\begin{abstract}
\noindent Steady-state and dynamics of the self-phase-locked
$(3\omega\longrightarrow2\omega, \omega)$ subharmonic optical
parametric oscillator are analyzed in the pump-and-signal resonant
configurations, using an approximate analytical model and a full
propagation model. The upper branch solutions are found always
stable, regardless of the degree of pump enhancement. The domain
of existence of stationary states is found to critically depend on
the phase-mismatch of the competing second-harmonic process.
\end{abstract}

\maketitle

\noindent A new class of subharmonic (frequency divide-by-three,
or $3\div1$) optical parametric oscillators (OPOs), namely
\emph{self-phase-locked} (SPL-OPOs), has aroused lately much
interest from both the experimental \cite{boller,kobayashi} and
theoretical \cite{zondyspl,longhipre,longhieur} standpoints.
Consider a $3\div1$ OPO pumped at an angular frequency
$\omega_{p}=3\omega$ (generating signal and idler waves at
$\omega_{s}=2\omega$ and $\omega_{i}=\omega$) containing a second
nonlinear crystal that is phase-matched for the competing
second-harmonic (degenerate down-conversion) of the idler (signal)
waves. The loss-free $\chi^{(2)}$ medium of total length
$L=L_{1}+L_{2}$ consists of a dual-grating periodically-poled (PP)
crystal comprising a first section of length $L_{1}$ perfectly
phase-matched for the
$\chi^{(2)}_{\text{OPO}}\equiv\chi^{(2)}(-3\omega;2\omega,\omega)$
down-conversion, followed by a second section of length $L_{2}$
nearly phase-matched for the competing
$\chi^{(2)}_{\text{SHG}}\equiv\chi^{(2)}(-2\omega;\omega,\omega)$
process with a wavevector mismatch $\Delta
k=k_{2\omega}-2k_{\omega}\neq 0$. Due to the mutual self-injection
of the subharmonic waves, the dynamics of the signal-and-idler
resonant devices was recently shown to depart from that of a
conventional ($L_{2}=0$) non-degenerate OPO \cite{zondyspl}. The
main difference is that SPL-OPOs are characterized by an intensity
bistability (sub-critical bifurcation) whatever the
configuration~\cite{zondyspl,longhipre,longhieur}. Secondly, each
intensity state of the stable branch can take 3 possible
deterministic phase values equally spaced by $2\pi/3$, while
conventional OPOs are subject to a stochastic phase diffusion
process~\cite{Graham}. Experimentally, $3\div1$ SPL-OPOs are
investigated in frequency
metrology~\cite{boller,douilletsyd,ikegami}, Fourier synthesis of
attosecond pulse \cite{kobayashi}, transverse pattern
formation~\cite{longhipre,longhieur} and potentially for new
features in squeezed states of light. Finally, owing to the much
lower pump intensity requirement, SPL-OPOs should be also suited
for the first experimental evidence of a Hopf bifurcation in cw
OPOs, predicted to only occur in triply-resonant OPOs under
extreme detuning conditions~\cite{lugiato}.

In this brief report, I provide an analysis of the pump-enhanced
singly resonant device (SPL-PRSRO) which was only over-viewed in
the conclusion of Ref.~\cite{zondyspl}. Bearing in mind that
PRSROs are easier to implement than DRO/TROs (subject to mode pair
instabilities) or SROs (requirement of high pump thresholds), it
is interesting to check how the dynamics of the doubly/triply
resonant SPL-OPOs is affected when only one of the subharmonic
waves oscillates with a varying pump enhancement, and to compare
with the behaviour of the purely idler-resonant (SPL-IRO) case
treated by Longhi in the mean-field approximation and neglecting
pump depletion~\cite{longhieur}.

The analysis starts with the solutions of the reduced propagation
equations for the normalized slowly varying field envelopes
$A_{j}(t',Z)=g_{1}L_{1}N_{j}(t',Z)$ ($j=p,s,i$) throughout the
dual-section medium. The complex amplitudes are scaled such that
their intensities $I_{j}=|A_{j}|^{2}$ are proportional to the
number of photons $|N_{j}|^{2}$ in $j$-th mode, times the square
of the small-signal parametric gain $g_{1}L_{1}$ ($\ll 1$) of the
$3\div1$ process ($g_{1}\propto
\chi^{(2)}_{\text{OPO}}$)~\cite{zondyspl}. In the phase-retarded
time frame ($t=t'-\bar{n}Z/c$; $z=Z/L_{1}$ or $z'= Z/L_{2}$),
where $\bar{n}$ is the average index of refraction, the 3
plane-wave equations for $0\leq z\leq L_{1}/L_{1}=1$ are
\begin{equation}
\partial_{z} A_{p}=iA_{i}A_{s} \,;\qquad \partial_{z} A_{s,i}=iA_{p}A_{i,s}^{\ast}, \label{eq:subpart-opo}
\end{equation}
with the initial condition $A_{i}(z=0)=0$ (non-resonant idler). In
the SHG section ($0\leq z'\leq L_{2}/L_{2}=1$), the subharmonic
amplitudes evolve as
\begin{eqnarray}
\partial_{z'} A_{s}&=&iSA_{i}^{2}\exp{(+i2\xi z')}, \label{eq:sigpart-shg} \\
\partial_{z'} A_{i}&=&iSA_{s}A_{i}^{\ast}\exp{(-i2\xi z')} \label{eq:idlpart-shg}
\end{eqnarray}
with the initial conditions $A_{s,i}(z'=0)=A_{s,i}(z=1)$, while
the pump amplitude keeps its value at $z=1$ throughout the second
section ($\partial_{z'} A_{p}=0$). The parameter $\xi=\Delta k
L_{2}/2$ is the phase mismatch of the competing SHG process. The
nonlinear coupling parameter $S=g_{2}L_{2}/g_{1}L_{1}$ is the
ratio of the SHG to OPO small signal gains ($g_{2}\propto
\chi^{(2)}_{\text{SHG}}$). Its expression reduces to $S\simeq
(L_{2}/L_{1})/\sqrt{3}$ for a $3\div 1$ OPO employing a PP
material~\cite{zondyspl}. The time-dependent cavity dynamics is
obtained from an iterative mapping of the resonating field
amplitudes at $z'=1$ and at a time $t$ to their values at $z=0$
after one roundtrip time $\tau$ of the ring-type cavity,
\begin{eqnarray}
A_{j}(t+\tau ,z=0) &=&r_{j}\exp (i\Delta
_{j})A_{j}(t,z'=1)+A_{in}, \label{eq:bound}
\end{eqnarray}
with $j=p,s$ and with $r_{j}$ being the (real) amplitude
reflectivity from $z'=1$ back to $z=0$. The constant input field
$A_{in}$ stands for the driving pump field and is null for the
signal wave. The $r_{j}$'s are related to the cavity loss
parameters $\kappa_{j}$'s by $\kappa_{j}=1-r_{j}$. For the
resonating subharmonic, it will be always assumed that
$\kappa_{s}\ll 1$, $r_{s}\simeq 1$. The amplitude loss parameter
$\kappa_{s}$ is then related to the cavity finesse by
$\textrm{F}_{s}=\pi/\kappa_{s}$ and to the cavity half linewith
$\gamma_{s}$ by $2\pi\gamma_{s}=\kappa_{s}/\tau$. The phase
factors $\Delta_{j}$ correspond to the linear propagation (and
mirror) phase shifts, modulo $2\pi$. The $\Delta$'s, also called
cavity detuning parameters, are equal to $(\nu-\nu_{c})\tau$, e.g.
to the wave frequency mismatches from the nearest cold cavity
frequency, scaled to the free spectral range $1/\tau$.

The dynamics of the systems can be numerically studied without any
approximation by solving
Eqs.(\ref{eq:subpart-opo}-\ref{eq:idlpart-shg}) using a
fourth-order Runge-Kutta solver with the appropriate initial
conditions and making use of the boundary conditions
(\ref{eq:bound}) (\emph{propagation} model). Unstable fixed point
cannot be found numerically so that approximate analytic solutions
of the steady state equations must be worked out by expanding the
amplitudes in Mac Laurin series of $z$, e.g.
$A_{j}(z)=A_{j}(0)+\sum_{n=1}^{\infty}[\partial_{z}^{(n)}A_{j}]_{0}\,z^{n}/n!$,
which allows to integrate
Eqs.(\ref{eq:subpart-opo}-\ref{eq:idlpart-shg}). Such an expansion
is justified by the smallness of the scaled amplitudes
($|A_{j}|\leq 1 $) since $g_{1}L_{1}\ll1$ and because $ z,z'< 1$.
The $n$-th order derivatives can be evaluated in terms of the
field products at $z=0$ using the generic equations
(\ref{eq:subpart-opo})-(\ref{eq:idlpart-shg}). %The number of terms
%kept in the field expansions must be consistent with the desired
%final order of the nonlinear coupling terms after the two
%successive $z$-integrations.
To get the leading(fourth)-order
coupling terms from this perturbative approach, only the $n=1$
terms in the field expansions need to be kept. After some algebra,
the approximate solutions of
(\ref{eq:subpart-opo})-(\ref{eq:idlpart-shg}) are
\begin{eqnarray}
A_{p}(t,L_{1}+L_{2}) &=&A_{p}(t,0)-\text{\footnotesize{(1/2)}}
A_{p}(t,0)\left| A_{s}(t,0)\right| ^{2}, \label{eq:prop1}
\\ A_{s}(t,L_{1}+L_{2})
&=&A_{s}(t,0)+\text{\footnotesize{(1/2)}} A_{s}(t,0)\left|
A_{p}(t,0)\right| ^{2} \nonumber \\ &-&i\chi^{\ast
}A_{p}^{2}(t,0)[A_{s}^{\ast }(t,0)]^{2}, \label{eq:prop2}
\\ A_{i}(t,L_{1}+L_{2})
&=&iA_{p}(t,0)A_{s}^{\ast }(t,0)+\chi A_{p}^{\ast
}(t,0)A_{s}^{2}(t,0). \label{eq:prop3}
\end{eqnarray}
where the nonlinear coupling parameter is
$\chi=S\exp{(-i\xi)}(\sin{\xi}/\xi)$. These solutions which assume
a linear $z$-variation of the fields depart from the numerical
ones for decreasing pump resonance (SRO limit), but still account
for pump depletion to first order. The cubic term in
Eq.(\ref{eq:prop1}), present in conventional PRSRO model
\cite{schiller}, is due to the usual cascading
$(3\omega-2\omega=\omega)$ followed by the re-combination process
$(\omega+2\omega=3\omega)$, while the last quartic term in
Eq.(\ref{eq:prop2}), describing the two-step processes
$(3\omega-2\omega=\omega)$ followed by $(\omega+\omega=2\omega)$,
leads to injection-locking. Note that this term is quadratic in
the signal-and-idler resonant cases (see Eqs.(9) in
Ref.~\cite{zondyspl}) or in the pure IRO case without pump
depletion (see Eq.(11) of Ref.~\cite{longhieur}). Stationary
solutions to Eqs.(\ref{eq:bound})-(\ref{eq:prop2}) are obtained by
requiring that $A_{p,s}(t+\tau,0)= A_{p,s}(t,0)\equiv A_{p,s}$.
Considering small enough detuning ($\Delta_{p,s}\ll 2\pi$), the
exponential phase factor in Eq.(\ref{eq:bound}) is expanded as
$\sim 1+i\Delta_{j}$. The resulting steady state amplitude
equations are
\begin{eqnarray}
(\kappa _{p}-ir_{p}\Delta _{p})A_{p} &=&-\text{\footnotesize{(1/2)}}
r_p(1+i\Delta _{p})A_{p}\left| A_{s}\right| ^{2}+A_{in}, \nonumber%\label{eq:stat1}
\\\frac{\kappa _{s}-i\Delta _{s}}{r_{s}(1+i \Delta_{s})}A_{s}
&=&\left[+\text{\footnotesize{(1/2)}} A_{s} A_{p}\right|
^{2}-i\chi ^{\ast }A_{p}^{2}[A_{s}^{\ast }]^{2}. \label{eq:stat2}
\end{eqnarray}
The above Mac-Laurin solutions (as compared with the mean-field
approach based on amplitude expansion in the power of the
$\kappa$'s~\cite{longhieur}) converge satisfactorily to the full
propagation model as long as $r_{p}\geq 0.8$. Excellent
convergence (to $\pm3\%$ for the parameters of
Fig.\ref{fig:figbr3}, for instance) is found when both the pump
and signal experience low roundtrip loss, not exceeding a few
percent even for large pumping ($I_{in}=|A_{in}|^{2}$ up to
$\sim$50 times the threshold for oscillation). For decreasing pump
resonance and moderate pumping they still provide a qualitative
account of the dynamical behavior of the system, although
resulting in higher values of the intensities $I_{j}=|A_{j}|^{2}$.
The domain of validity of the Mc Laurin model will be shown to
depend on the value of $r_{p}$. In the true SRO limit
($r_{p}\rightarrow 0$), the full propagation model remains always
valid.

Besides the trivial (non-lasing) solutions $A_{s,i}=0$,
 $A_{p}=A_{in}/(\kappa _{p}-ir_{p}\Delta _{p})$,
Eqs.(\ref{eq:stat2}) admit non-zero intensity states
$I_{p,s}=|A_{p,s}|^{2}$ with well-defined phases. It is convenient
to introduce the scaled intensities
$\bar{I}_{p}=I_{p}/I_{p}^{th}$, $\bar{I}_{s}=I_{s}/I_{S}$,
$\bar{I}_{i}=I_{i}/(I_{p}^{th}I_{S})$, with $I_{p}^{th}=2\kappa
_{s}$, $I_{S}=2\kappa _{p}/r_{p}$; input pump intensity
$\bar{I}_{in}=I_{in}/(2\kappa_{p}^{2}\kappa_{s})$ ; and normalized
cavity detuning $\bar{\Delta}_{p,s}=\Delta_{p,s}/\kappa_{p,s}$.
Introducing $C_{s}=4\left| \chi \right| ^{2}(\kappa _{p}/r_{p})$,
and taking the modulii of (\ref{eq:stat2}), the scaled signal
intensity $\bar{I}_{s}$ is then the solution of
\begin{equation}
\begin{split}
\left[F(1-r_{s}\kappa_{s}\bar{\Delta}_{s}^{2})-r_{s}(1+\kappa_{s}^{2}\bar{\Delta}_{s}^{2})\bar{I}_{in}\right]^{2}+
(\bar{\Delta}_{s}F)^{2} \\
 = 2\,C_{s}\,%\stackrel{}{(\kappa_{p}/r_{p})
 r_{s}^{2}(1+\kappa_{s}^{2}\bar{\Delta}_{s}^{2})^{2}\bar{I}_{s}\,\bar{I}_{in}^{2},
\end{split}
 \label{eq:statsca1}
\end{equation}
where the symbol $F$ stands for
\begin{equation}
F = (1+\bar{I}_{s})^{2}+\bar{\Delta}_{p}^{2}\,(\kappa _{p}\bar{I}%
_{s}-r_{p})^{2}. \label{eq:statsca3}
\end{equation}
The intracavity pump is given by $F\bar{I}_{p}=\bar{I}_{in}$ and the idler
intensity by
$\bar{I}_{i}=\bar{I}_{p}\bar{I}_{s}[1+\bar{I}_{S}|\chi|\bar{I}_{s}+(\bar{I}_{p}-1)/\bar{I}_{p}]$.

Phase relationships, demonstrating phase-locking of the
subharmonic waves to the pump laser, can be derived from
Eqs.(\ref{eq:stat2}) by writing
$A_{j}=\alpha_{j}\exp{(i\varphi_{j})}$ where $\alpha_{j}$ are the
amplitude modulii. Defining $\varphi_{in}$ as the arbitrary phase
of the pump laser and $\varphi_{D}=\xi+2\varphi_{p}-3\varphi_{s}$,
one obtains
\begin{eqnarray}
%\tan{(\varphi_{in}-\varphi_{p})}&=&-\frac{\bar{\Delta}_{p}(1-\kappa_{p}\bar{I}_{s})}{1+\bar{I}_{s}},%
%\label{eq:tanphip} \\ \cot{(\xi+2\varphi_{p}-3\varphi_{s})}&=&
%\frac{\bar{\Delta}_{s}}{1-\bar{I}_{p}} \label{eq:cotanphi2}
\tan{(\varphi_{in}-\varphi_{p})}= -[\bar{\Delta}_{p}(r_{p}-\kappa_{p}\bar{I}_{s})]/[1+\bar{I}_{s}],%
\label{eq:tanphip} \\
\cot{\varphi_{D}}=
\bar{\Delta}_{s}/[1-r_{s}\kappa_{s}\bar{\Delta}_{s}^{2}-r_{s}(1+\kappa_{s}^{2}\bar{\Delta}_{s}^{2})\bar{I}_{p}].
\label{eq:cotanphi2}
\end{eqnarray}
When solved for $\varphi_{s}$, these relations yield 3 possible
values $\varphi_{s}=\varphi_{0}+2k\pi/3$ ($\varphi_{0}$ being a
constant and $k=0,\pm1$), while for $|\chi|=0$ only the sum phase
$\varphi_{s}+\varphi_{i}=\varphi_{p}+\pi/2$ is deterministic as
predicted for conventional oscillators, due to the phase diffusion
noise stemming from the spontaneous parametric fluorescence
~\cite{Graham}. Furthermore, when $|\chi|=0$ ($C_{s}=0$),
Eq.(\ref{eq:statsca1})
implies that the signal resonates necessarily with zero detuning ($\bar{\Delta}_{s}=0$%
), and one retrieves the result that the intracavity pump is
 clamped to the constant value $\bar{I}_{p}=1$ for any $\bar{I}_{in}$ \cite{schiller}.
\begin{figure} [htb]
    \begin{center}
        \includegraphics [height=2.5cm]{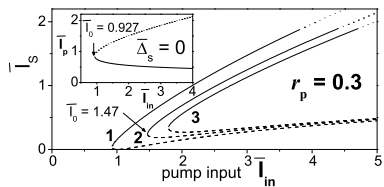}
    \end{center}
    \caption{Bifurcation diagram of signal intensity versus pump
    parameter, computed from the Mc Laurin solutions, for $r_{p}=0.3$,
    $\bar{\Delta_{p}}=0$, $\kappa_{s}=0.005$, $\chi=S=0.2$. Curves
    (1)-(3) are for $\bar{\Delta}_{s}=0; 0.4; 0.5$. The inset plots
    show the intracavity pump stable (solid line) and unstable
    (dashed) fixed points $\bar{I}_{p}^{\pm}$. Note that
    $\bar{I}_{s}^{+}$ diverges at $\bar{I}_{in}\geq 3.8;4.4;4.7$ for
    curves 1;2;3 from either the LSA analysis or the time mapping of
    Eqs.(\ref{eq:prop1})-(\ref{eq:prop3}), setting the validity range
    of the Mc Laurin approximation.} \label{fig:figbr1}
\end{figure}
From Eq.(\ref{eq:statsca3}), $\bar{I}_{s}$ is then the solution of
a quadratic equation which admits a single positive solution
(supercritical bifurcation) if and only if $\bar{I}_{in}(1+\kappa
_{p}^{2}\bar{\Delta}_{p}^{2})-\bar{\Delta}_{p}^{2}\geq 0$. This
condition is always satisfied for $\bar{I}_{in}\geq \bar{I}_{th}$,
where the input (intracavity) pump threshold expresses as
$\bar{I}_{th}=1+r_{p}^{2}\bar{\Delta}_{p}^{2}$. Considering now
the case $|\chi|\neq 0$, the signal wave is \emph{a priori} no
longer constrained to oscillate with zero detuning and its
intensity is the solution of a quartic equation,
$\sum_{n=0}^{4}a_{n}\bar{I}_{s}^{n}=0$, obtained by expanding
Eq.(\ref{eq:statsca1}) using Eq.(\ref{eq:statsca3}).
%Closed-form solutions cannot be worked out as in the case of the
%doubly resonant~\cite{zondyspl} or singly
%resonant~\cite{longhieur} devices, so that an analytic formulation
%of the boundary of their existence is precluded.
 The numerical resolution of this equation for a wide range of
signal detuning or driving pump intensity always yield two real
positive roots $\bar{I}_{s}^{\pm}$, defining two branches of
solutions, for pump intensities $\bar{I}_{in}\geq \bar{I}_{0}$
(see below for definition of $\bar{I}_{0}$). The stability of
these two fixed points, each associated with the 3 possible phase
states, was investigated using a linear stability analysis
(LSA)~\cite{zondyspl} that leads to a quartic characteristic
equation, $\sum_{n=0}^{4}\Phi_{n}\Lambda^{n}=0$. The LSA results
were double-checked by a direct time mapping of
Eqs.(\ref{eq:prop1})-(\ref{eq:prop2}) using the boundary
conditions (\ref{eq:bound}). From the LSA of the trivial state,
the threshold for oscillation (not necessarily on a stable fixed
point) remains the same as for conventional devices.
Fig.\ref{fig:figbr1} shows the analytical bifurcation plot versus
the pump parameter for $r_{p}=0.3$ and 3 different values for
$\bar{\Delta}_{s}$, as compared with the solutions computed with
the propagation model (Fig.\ref{fig:figbr2}). The bifurcation
diagram versus $\bar{I}_{in}$ displays a saddle-node region, with
$\bar{I}_{0}$ being the input intensity at the saddle-node point
where $\bar{I}_{s}^{+}=\bar{I}_{s}^{-}$. For zero or vanishingly
small $\bar{\Delta}_{s}$, one has $\bar{I}_{0}<\bar{I}_{th}$ but
as the detuning is increased, $\bar{I}_{0}>\bar{I}_{th}$. This
behavior contrasts with the signal-and-idler resonant cases, for
which a transition from sub-criticality to super-criticality is
predicted when $\bar{\Delta}_{s,i}\rightarrow0$, corresponding to
the merging of the saddle-node point of coordinate
($\bar{I}_{0},\bar{I}_{s}\neq0$) with the threshold point
($\bar{I}_{th},\bar{I}_{s}=0$)~\cite{zondyspl}.
\begin{figure} [htb]
        \begin{center}
            \includegraphics [height=2.5cm]{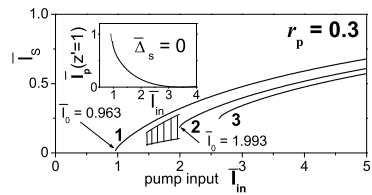}
        \end{center}
    \caption{Stable (upper) branches of signal intensity versus pump
    parameter, computed from the propagation model with the same
    parameters as for Fig.\ref{fig:figbr1}. The hatched domain at the
    left of curve (2) gives the amplitude of the limit cycles below
    the saddle-node intensity $\bar{I}_{0}$ (non-stationary SPL
    states). The inset plots show the output ($z'=1$) intracavity pump
    $\bar{I}_{p}$.} \label{fig:figbr2}
\end{figure}
Note also the tiny range of sub-threshold states in curve (1),
resulting from the weak self-injection regime. From the LSA of the
Mc Laurin solutions, the $\bar{I}_{s}^{-}$ branches (dashed
curves) are always unstable, due to a positive real eigenvalue,
but only a portion of the upper branch (solid lines) extending
from the saddle-node intensity $\bar{I}_{0}$ to some critical
intensity $\bar{I}_{C}$ is found stable. The instability (dotted
lines) beyond $\bar{I}_{C}$ (characterized by 3 real positive
eigenvalues of the LSA equation) was confirmed by the time mapping
of the Mc Laurin solutions which diverges at $\bar{I}_{C}$,
wherewhile the full propagation model converge to a fixed point
for any input intensity value (Fig.~\ref{fig:figbr2}). Hence the
whole upper branch of the SPL-PRSRO is actually stable, the
instability predicted by the Mc Laurin model being merely an
artefact of the approximation. Actually, as the pump enhancement
is decreased the validity range of the Mc Laurin model is
restricted to pump parameters lying closer and closer to the
threshold. In Fig.~\ref{fig:figbr2}, obtained by backward
adiabatic following of the stationary solutions, all curves end at
their saddle-node point $\bar{I}_{0}$ since critical slowing down
is observed as these points are approached. The limit cycles
occurring for $\bar{I}_{in}<\bar{I}_{0}$ (hatched area) are merely
due to the non-existence of stationary states. The upper branch is
found stable even in the SRO limit ($r_{p}=0$), whatever the
detunings, in contrast with the SPL-IRO mean-field analysis
results~\cite{longhieur}. Let us remind that in the SPL-DRO/TROs
the upper branch was found to destabilize via a Hopf
bifurcation~\cite{zondyspl}. The difference in dynamical behavior
is due to the stronger self-injection regime in these latter
devices. The inset frame in Fig.~\ref{fig:figbr2} gives the output
pump intensity (at $z'=1$) corresponding to $\bar{\Delta}_{p,s}=0$
(curve (1)), it can be seen that the pump is no longer clamped to
unity as in conventional PRSROs~\cite{schiller}, meaning that the
competing nonlinearity enhances the down-conversion efficiency.
\begin{figure} [htb]
    \begin{center}
        \includegraphics [height=2.5cm]{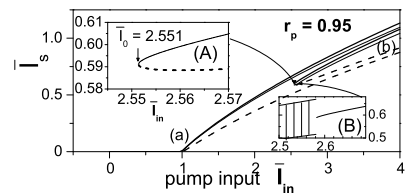}
    \end{center}
    \caption{Bifurcation diagram of signal intensity versus pump
    parameter, computed from the analytical model for $r_{p}=0.95$ and
    same other parameters as in Fig.\ref{fig:figbr1}, excepted that
    curves (a)-(b) are for $\bar{\Delta}_{s}=0; 0.1$. The thin solid
    lines under-riding the (a)-(b) solid lines show the upper branch
    computed from the propagation model. The insets (A: Mc Laurin
    model, B: propagation model) are blow-ups of the saddle-node
    region of (b) case. In (B), critical slowing-down characteristic
    of saddle-node points occurs.} \label{fig:figbr3}
\end{figure}
\begin{figure} [htb]
    \begin{center}
        \includegraphics [height=2.5cm]{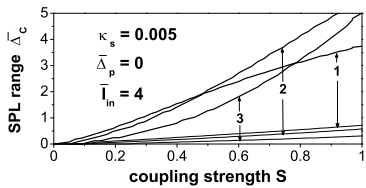}
    \end{center}
    \caption{Critical values of the signal detuning parameter
    delimiting the boundary of stable (underneath the curves) and
    oscillatory SPL states (above) in the ($\bar{\Delta}_{s},S$) plane
    at fixed pump input/detuning, computed from the propagation model.
    The thick solid lines are for $r_{p}=0.3$ and the thin lines for
    $r_{p}=0.95$. The curve labels (1)-(3) stand for
    $\xi=0;\,\pi/2,\,3\pi/4$.} \label{fig:figbr4}
\end{figure}
In Fig.~\ref{fig:figbr3}, as the pump finesse increases
($\kappa_{p}=0.05$), the Mc-Laurin (thick solid lines) and the
propagation model (thin solid lines) converge excellently for any
input intensity. Both models predict then a whole upper branch
stability, for a wide range of input intensity (up to tested
$\bar{I}_{in}=50$ at least). The Mc Laurin model converges to the
propagation model because when $\kappa_{p}\rightarrow 0$ the
assumption of a linear $z$-dependence of the resonating fields
inside the medium is fully justified (uniform field limit). Notice
that the threshold value is then more sensitive to the signal
detuning than in Fig.~\ref{fig:figbr2}. As a consequence of the
reduced sub-threshold state range due to the weak SPL regime one
expect that the self-locking detuning range, defined as the
maximum allowed $\bar{\Delta_{s}}$ for a given pump input
intensity, would be smaller than in signal-and-idler resonant
set-ups. Indeed large pump enhancement is paid back with a
shrinking self-locking range. Fig.\ref{fig:figbr4} displays the
critical $\bar{\Delta}_{C}(S)$ detuning for both $r_{p}=0.3$ and
$r_{p}=0.95$, when $\bar{\Delta}_{p}=0$ and when the device is
pumped 4 times above threshold. Surprisingly, for the same
nonlinear coupling strength $S$, the high pump resonance case
leads to less than a cavity linewidth SPL range, while for the SRO
limiting case the SPL range is 10-fold wider, as with
SPL-DRO/TROs.
\begin{figure} [t]
    \begin{center}
        \includegraphics [height=3cm]{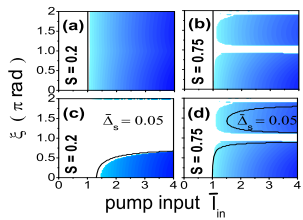}
    \end{center}
    \caption{Domain of stability (dark region) of phase-locked states
    in the $(\xi, \bar{I}_{in})$ plane (propagation model), with
    $r_{p}=0.95$, $\kappa_{s}=0.005$, $\bar{\Delta}_{p}=0$: (a),(c)
    for $S=0.2$ and (b),(d) for $S=0.75$. Upper (lower) frames are for
    $\bar{\Delta}_{s}=0$ ($\bar{\Delta}_{s}=0.05$). The solid lines
    give the boundary of existence of stationary states from the Mc
    Laurin model. The plots are symmetric for negative phase mismatch
    $\xi$.} \label{fig:figbr5}
\end{figure}
Hence a large SPL range is not necessarily associated with the
double-resonance condition that involves a strong mutual-injection
process between the subharmonics. The small ($<1$-MHz, e.g. less
than a cavity linewidth) self-locking range reported by Boller
\emph{et al} \cite{boller} for a SPL-PRSRO is in agreement with
this expectation. As a noticeable difference with the
signal-and-idler resonant cases, the coefficients of the LSA
equation depend on the phase of the coupling parameter $\chi$,
while $\chi$ enters only as its modulus in the SPL-DRO stability
analysis (see the Appendix in Ref.\cite{zondyspl}). One hence
expects that the dynamics of SPL-PRSROs will be sensitive to the
SHG phase mismatch $\xi$. Fig.\ref{fig:figbr5} shows the
steady-state signal intensity contour plots in the parameter space
($\bar{I}_{in},\xi$), computed with the propagation model for
$\bar{I}_{in}=4$, $r_{p}=0.95$, $\bar{\Delta}_{p}=0$ and two
values $S=0.2$ (a,c) and $S=0.75$ (b,d) of the coupling parameter.
For the stronger coupling $S=0.75$ (panels b-d), small amplitude
limit cycles arise around $\xi=\pm \pi$, even for
$\bar{\delta}_{s}=0$ (panel b). To avoid such non-stationary
states, it is important to tailor accurately the simultaneous
phase-matching of both competing processes.

In conclusion, it is found that the whole upper branch of the
SPL-(PR)SRO is stable and that the self-locking range shrinks with
increasing pump enhancement. In the SRO limit, the approximated
model may fail to describe correctly the dynamical behavior over a
large pump input range. Of more concern, in contrast with
signal-and-idler resonant devices, the domain of existence of
stationary phase-locked states is found sensitive to the value of
the residual phase mismatch of the competing SHG nonlinearity. The
author is indebted to one of the referees for his personal
involvement in the improvement of this report. This work has
benefited from a partial support from an European Union
INCO-Copernicus grant (Contract No. ERBIC15CT980814).


\begin{thebibliography}{}
\bibitem{boller} K. Boller \emph{et al}, Opt. Expr. {\bf 5}, 114 (1999).
\bibitem{kobayashi} Y. Kobayashi, K. Torizuka, Opt. Lett. {\bf 25},
856 (2000).
\bibitem{zondyspl} J.-J. Zondy \emph{et al},
Phys. Rev. A {\bf 63}, 023814 (2001).
% See also CLEO/IQEC 2000 digest, paper no. QWA3, p.131, (2001).
\bibitem{longhipre} S. Longhi, Phys. Rev. E {\bf63}, 055202(R) (2001).
\bibitem{longhieur} S. Longhi, Eur. Phys. J.D {\bf17}, 57 (2001).
\bibitem{Graham}  R.\ Graham, H.\ Haken, Zeit. f\"{u}r Phys. {\bf 210},
276 (1968).
\bibitem{douilletsyd} A. Douillet \emph{et al}, IEEE Trans. Instr. Meas. {\bf 50}, 548 (2001).
\bibitem{ikegami} S. Slyusarev \emph{et al}, Opt. Lett. {\bf 24},
1856 (1999).
\bibitem{lugiato}  L. A. Lugiato et \emph{et al}, Il Nuovo Cimento {\bf 10D}, 959 (1988).
%\bibitem{Meanf}  M. Le Berre, A.S. Patrascu, E Ressayre, A. Tallet and
%N.I.\ Zheleznykh, Chaos, Solitons, \& Fractals {\bf 4}, 1389 (1994).
\bibitem{schiller} S. Schiller \emph{et al}, \josab
 {\bf 16}, 1512-1524 (1999).
\end{thebibliography}
\end{document}